\documentclass[preprint]{aastex}
\usepackage{emulateapj5,epsfig}

\newcommand{\bu}{SN~1998bu}

\newcommand{\Msun}{M$_{\odot}$}

\shorttitle{}
\shortauthors{Cappellaro et al.}

\begin{document}


\title{Detection of a light echo from SN~1998\lowercase{bu}\altaffilmark{0}.}


\author{E. Cappellaro\altaffilmark{1}, F. Patat\altaffilmark{2},  
P. A. Mazzali\altaffilmark{3}}
\author{ S. Benetti\altaffilmark{1}, 
J.I. Danziger\altaffilmark{3}, A. Pastorello\altaffilmark{4},
L. Rizzi\altaffilmark{1,4},
M. Salvo\altaffilmark{4}, M. Turatto\altaffilmark{1}}

\altaffiltext{0}{Based on observations obtained at ESO-La Silla}
\altaffiltext{1}{Osservatorio Astronomico di Padova, vicolo dell'Osservatorio, 5, 
Padova, Italy }
\altaffiltext{2}{European Southern Observatory, Karl-Schwarzschild-Strasse 2,
D-85748 Garching bei M{\"u}nchen, Germany}
\altaffiltext{3}{Osservatorio Astronomico di Trieste, Via Tiepolo, 11, 
Trieste, Italy}
\altaffiltext{4}{Dipartimento di Astronomia, Universit\`a di Padova, 
vicolo dell'Osservatorio, 2, Padova, Italy}


\begin{abstract}
About 500d after explosion the light curve of the Type Ia \bu\/
suddenly flattened and at the same time the spectrum changed from the
typical nebular emission to a blue continuum with broad absorption and
emission features reminiscent of the SN spectrum at early phases. We
show that in analogy to SN~1991T \citep{schmidt}, this can be
explained by the emergence of a light echo from a foreground dust
cloud. Based on a simple model we argue that the amount of dust
required can consistently explain the extinction which has been
estimated by completely independent methods.  Because of the similar
echo luminosity but much higher optical depth of the dust in
\bu\/ compared with SN~1991T, we expect that the echo 
ring size of \bu\/ grows faster than in SN~1991T. HST observations have
indeed confirmed this prediction.
  
\end{abstract}




\section{Introduction} 

The Type Ia SN~1998bu was discovered on May 9.9 UT by \citet{villi} in
a spiral arm of NGC~3368 (M96), a nearby Sab galaxy.  It had 
very good photometric and spectroscopic coverage by several groups
\citep{jha,hernandez,suntzeff} showing that the luminosity decline was
slower than the average ($\Delta m_{15}(B)$, the magnitude decline
in the early 15 days after maximum, was $1.01\pm0.05$ mag). This is
almost as slow as the slowest SN~Ia on record, SN~1991T, although,
unlike SN~1991T, SN~1998bu was spectroscopically normal before and
around maximum.  The distance to the host galaxy was measured using
Cepheids variables \citep{tanvir} as $\mu=30.25\pm0.19$.  However, a
calibration of the absolute luminosity of the SN depends also on the
estimated reddening which in the case of SN~1998bu is
significant. \citet{hernandez} estimated $E(B-V) = 0.86\pm0.15$
based on the comparison of the spectral energy distribution of \bu\/
with that of the template SN~Ia 1981B.

\bu\/ was included in our program of monitoring of the nebular phase 
of SN~Ia. During this monitoring  we realized that \bu\/
was not declining with the expected rate; this, combined with the
change in the spectral appearance gave the first evidence of the
emergence of a light echo in \bu\/ \citep{capiau}.
These observations are presented here for the first time
along with a simple modeling.

\section{Evidence for a light echo in SN~1998\lowercase{bu}}

When the SN~Ia ejecta become optically thin, about 100 days after the
explosion, the SN luminosity is determined essentially by the
deposition of the kinetic energy of the positrons released by the
decay of $^{56}Co$.  There is evidence that even positron deposition
may not be complete
\citep{cap97,milne}, and so at advanced phases
the light curve declines either at the $^{56}Co$ rate ($0.98$
mag/100d) or faster. The only previously known case of a SN~Ia with a
decline slower than the $^{56}Co$ rate was SN~1991T, whose slow late
decline and peculiar spectra were interpreted as due to a light
echo
\citep{schmidt}.

A first indication that \bu\/ was beginning to deviate from the normal
exponential decline came on December 4 1999, roughly 500 days after
maximum.  At this epoch the observed magnitude ($V=20.75\pm0.08$) was
already 2 mag brighter than the extrapolation of the radioactive decay
tail. Additional photometric observations over the following few
months showed that the SN luminosity remained almost constant (the
decline was only $\sim 0.5$ mag in 100 days). Also, the color
at these epochs was unusually blue ($B-V \simeq -0.1$). This behavior
was reminiscent of that exhibited by SN~1991T \citep{schmidt,sparks}.

\begin{deluxetable}{llllllllll}
\tabletypesize{\scriptsize}
\tablecaption{SN~Ia data}
\tablewidth{0pt}
\tablehead{\colhead{SN} &
\colhead{galaxy} & 
\colhead{V(max)} & 
\colhead{A$_V^{gal}$} & 
\colhead{A$_V^T$} & 
\colhead{$\Delta m_{15}$(B)} &    
\colhead{Ref.} &
\colhead{$\mu$} &
\colhead{Ref.} }
\startdata
1991T  & NGC 4527 & $11.51\pm0.02$ & 0.07 & $0.53\pm0.17$ & $0.94\pm0.05$
       & \citet{phil99} & $30.74\pm0.12$ & \citet{saha} \\
1998bu & NGC 3368 & $11.88\pm0.02$ & 0.08 & $0.94\pm0.15$ & $1.01\pm0.05$
       & \citet{jha} & $30.25\pm0.19$ & Tanvir et al. (2000)\\ 
\enddata
\end{deluxetable}

Fig.~1 shows the absolute V light curves of \bu\/ and SN~1991T
\citep{cap97}. Both light curves have been calibrated using the
Cepheid-based distances \citep{saha,tanvir} and corrected for total
extinction (cf. Tab.~1).  The striking similarity of the two light
curves suggests that the same mechanism causing the flattening of the
light curve of SN~1991T was also acting in the case of \bu\/.

Although this is not clearly visible in Fig.~1, the absolute magnitude 
of \bu\/ at maximum was about 0.5 mag fainter than that of SN~1991T. 
Based on  the $\Delta m_{15}(B)$ vs $M_V$ relation
\citep{phil99} the difference should be only 0.05 mag. 
The gap increases slightly with time, and it reaches about 0.8-0.9 mag
one year after maximum.  That SNe~Ia with a very similar $\Delta
m_{15}(B)$ actually show significant dispersion in some of their other
properties is a well-known fact \citep{paolo98}, and is one of the
main caveats for the use of SN~Ia as distance indicators. However,
this is not essential for the discussion presented here.

{
\centerline{{\vbox{\epsfxsize=8cm\epsfbox{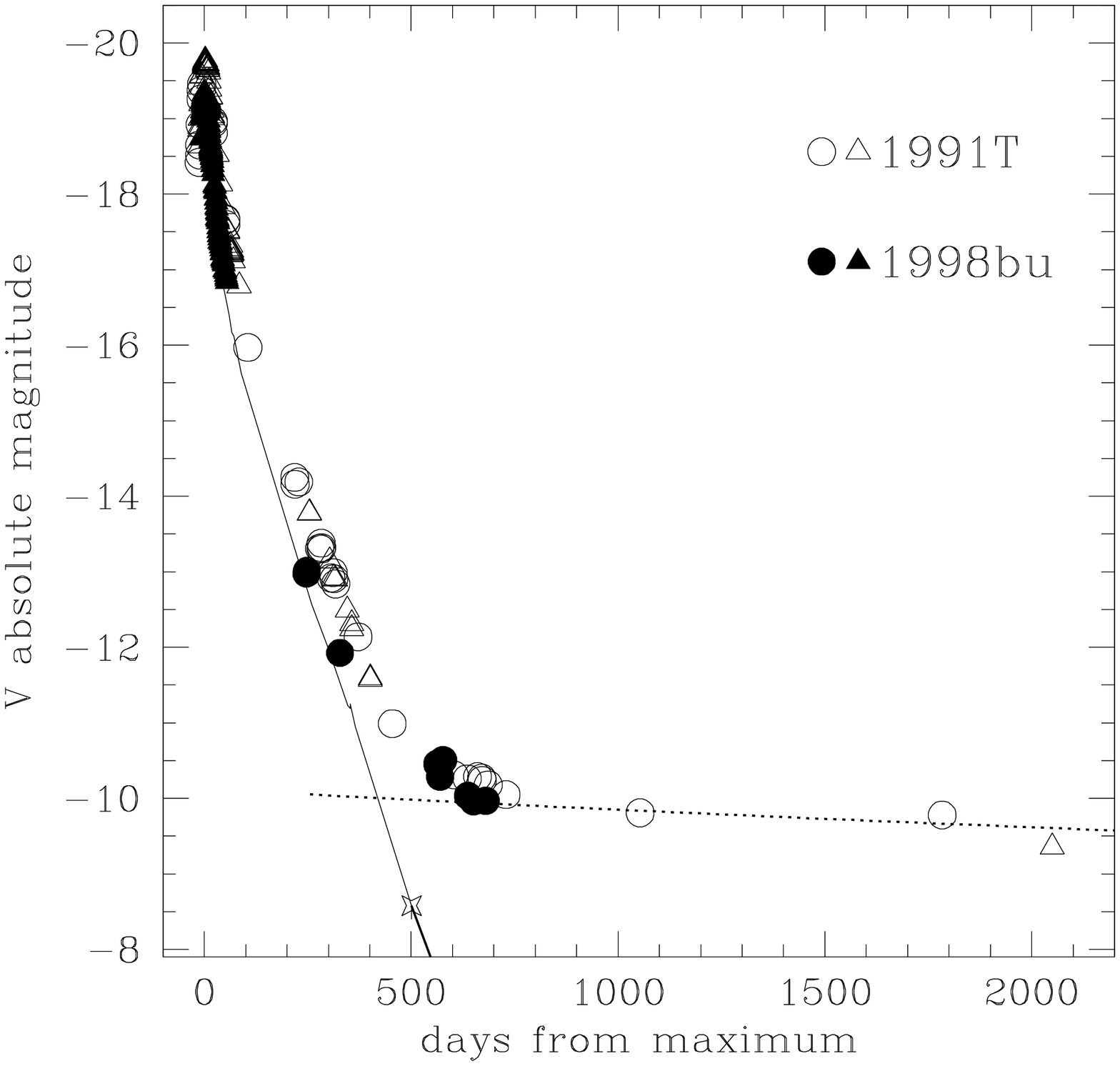}}}}
{\footnotesize{FIG. 1.  The light curve of SN 1998bu is compared with
that of SN~1991T (circles are our own measurements, whereas triangles
are data from literature).  For reference, the continuous line
represents the V absolute light curve of SN~1996X, a normally
declining SN~Ia (the star marks the epoch of the last measurement of
SN~1996X and after that the line is simply an extrapolation). The
dotted line is the fit of the observations with the simple model
described in Sect.~2.  }}
\bigskip
}

The slow photometric evolution of \bu\/ makes it a viable target
for spectroscopy even 2 yrs after explosion.  Spectra
of \bu\/ where obtained using the ESO~3.6m telescope (and EFOSC2) at La
Silla at two epochs, 2000 March 3 and March 30. We used grism
\#11 which, combined with the 1.2 arcsec slit, allowing one to cover the range
3300-7500 \AA\/ with a resolution of about 15\AA. Eventually, since no
spectral evolution was apparent between the two epochs, the two
spectra were merged to improve the S/N. The total integration time was
9000 sec.  The resulting spectrum, labeled with the average phase of
670d, is shown in Fig.~2. For comparison we also show the spectrum of
SN~1991T obtained with the same telescope at a similar phase. We also
show spectra of the two SNe at a much earlier epoch, when the
luminosity decline was still tracking the radioactive exponential
tail. 

The similarity of the spectral evolution of the two SNe is remarkable.  
About 300 days after maximum the spectra are dominated
by the strong emission lines of FeII] and FeIII] which are typical of
SNe~Ia in the nebular phase. These emission lines
originate in the iron nebula and are powered by the radioactive decay of
$^{56}$Co to $^{56}$Fe \citep{kuchner}. 

In correspondence to the flattening of the light curve the
spectra change completely, showing a blue continuum with superposed
broad absorption and emission features. These spectra resemble the
photospheric epoch spectra, but they do not match any specific
early-time epoch.  For SN~1991T it was convincingly suggested
\citep{schmidt} that the peculiar late time light curve and spectrum
can be explained assuming that the emission at these epochs is
dominated by the echo from circumstellar dust of the light emitted by
the SN near maximum. Our observations suggest that the
same mechanism is at work also in \bu\/.

The key element for such a mechanism to work is the presence of a dust
layer in the vicinity of the SN, scattering towards the observer a
fraction of the light emitted by the SN. Because of the different
travel times the observed spectrum is a combination of the early time
scattered spectra and of the direct, late-time spectrum.  In
principle, the details of the echo (intensity, duration, spectrum)
depend on the spatial distribution of the dust and on the physics of
the scattering process. A careful analysis and modeling of time
distributed observations may allow one to derive useful
constraints. Such a detailed study is the subject of future work
(Patat et al. in preparation), while here we illustrate the basic
principles of the phenomenon.

In general, the light curve of the echo is related to the light emitted
by the SN by the following relation \citep{chevalier}:

\begin{equation}\label{flux}
F_{echo}(t) = \int^t_0 \; F_{SN}(t-t^\prime) \;
f(t^\prime) \;dt^\prime
\end{equation}

where $F_{SN}(t-t^\prime)$ is the flux of the SN at time $t-t^\prime$
and $f(t)$, the fraction of this light which is scattered towards the
observer, depends on the geometry of the system and on the properties
of the dust (units of $f(t)$ are s$^{-1}$; cf. Equation~\ref{fdt}). As
a first approximation we can assume that the SN light curve is a short
pulse with a duration $\Delta t_{SN}$, during which the emitted flux
$F_{SN}$ is constant. Under this assumption Equation
\ref{flux} becomes $F_{echo}(t)= F_{SN} \;  f(t) \; \Delta t_{SN}$, where
$\Delta t_{SN}$ can be obtained by a numerical integration of the
observed light curve as $F_{SN} \Delta t_{SN} = \int_0^{+\infty}
F_{SN}(t) dt$. In the case of the $V$ light curve of SN~1991T
\citep{schmidt,sparks} this gives $\Delta t_{SN}$=0.11 yr.

As one of the possible geometries which are consistent with the
observations, we consider a thin dust sheet lying in front of the SN
and extended perpendicularly to the line of sight. For the sake of
simplicity we make the hypothesis that the sheet thickness $\Delta D$
is much smaller than the distance between the SN and the dust elements
and we consider only single scattering. Given this idealized
configuration and assuming the scattering phase function by Henyey \&
Greenstein (1941), an analytic expression for the function $f(t)$ can
be derived relatively easily \citep{chevalier,xu} as:

\begin{equation}\label{fdt}
f(t) = \frac{c}{8\pi}\,\frac{\omega_d \tau_d}{D+ct}\,
\frac{1-g^2}{(1+g^2-2g\frac{D}{D+ct})^{3/2}}
\end{equation}

where $D$ is the distance of the dust layer from the SN, $\tau_d$
is the optical depth of the dust layer, $\omega_d$ is the dust albedo,
and $g$ is a parameter which represents the degree of forward scattering.
The value of $g$ ranges from 0 for isotropic scattering to 1 for
purely forward scattering. Empirical estimates and numerical calculations
\citep{white} give $g\approx$0.6. For the dust albedo we have adopted 
$\omega_d\approx$0.6 \citep{mathis}.

{
\centerline{{\vbox{\epsfxsize=8cm\epsfbox{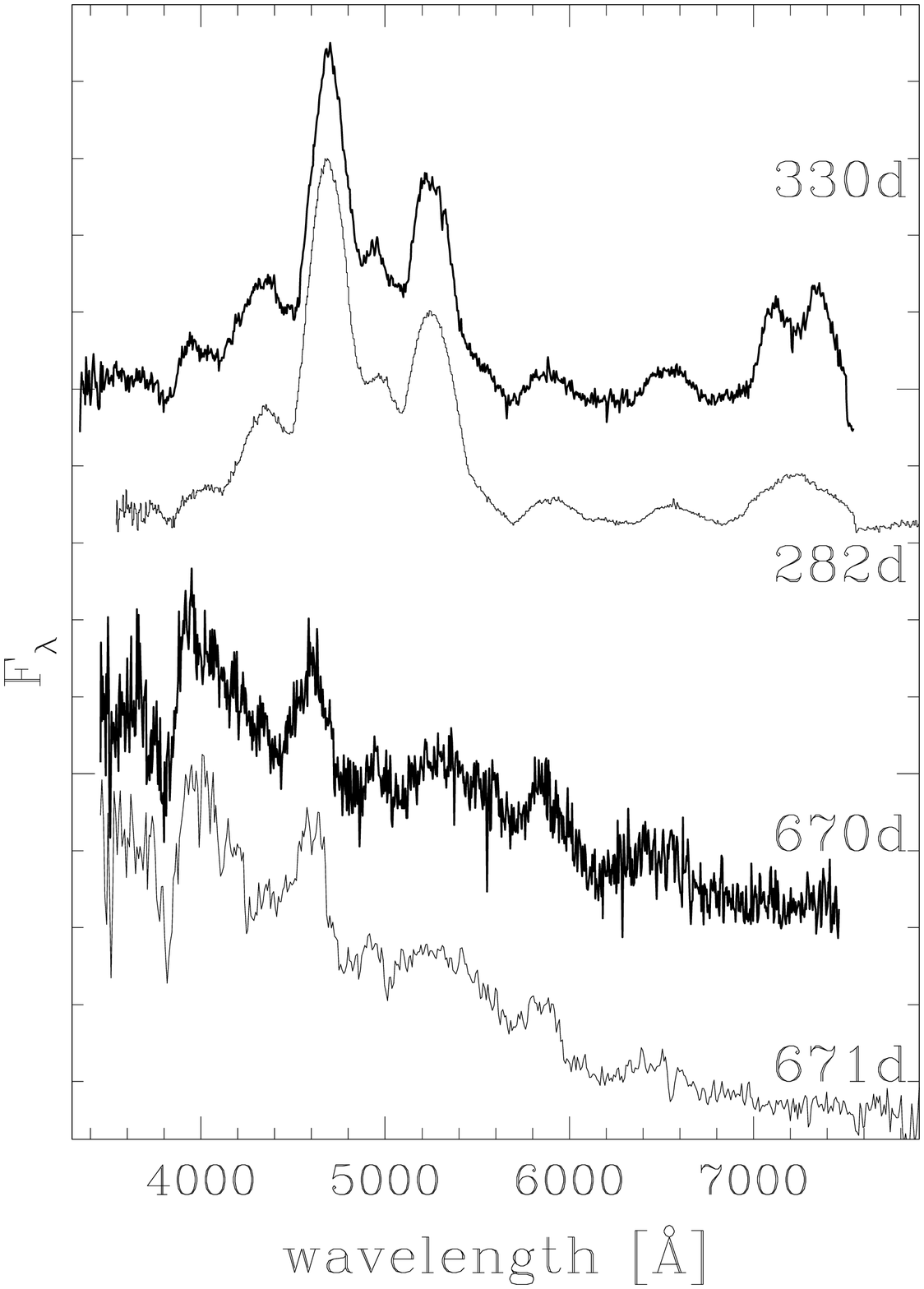}}}}
{\footnotesize{Comparison of the spectra of SN1998bu (thick lines) and SN1991T 
(thin lines) at two epochs, roughly 1 and 2 years past
explosion. Spectra have been reported in the host galaxy rest frames,
but have not been corrected for reddening}}
\bigskip}

Equation \ref{fdt} shows that if the dust is too close to the object,
even if it survives the exposure to the enormous radiation flux of the
SN, it would not produce a constant, long duration
echo. This is because of both the angular dependence of the scattering
phase function and the SN radiation dilution. In the case of SN~1991T the 
echo luminosity declined by a factor of 4 over $\sim 7$ yrs \citep{sparks} 
which implies $D > 45$ lyr (equation~\ref{fdt}).  On the other hand, if a 
dust layer is to produce a significant light echo, it must be located close 
enough to the SN that the emission from the SN is not too diluted. In fact, 
other conditions being similar, for a distant cloud ($D\gg ct$),
the echo flux is $F_{echo} \propto \tau_{d}/D$.  
In particular, since we found that the cloud must be located farther 
than 45 lyr from the SN, we derive a lower limit for the dust optical 
depth of $\tau_d>0.28$. Since in our model the dust layer is located in front 
of the SN  we expect that it causes some extinction on the line of sight. 
Assuming that the column density is constant along different lines of sight 
and based on the relation $A_V = 1.086\tau_V$ \citep{cox} we can estimate 
for SN~1991T a lower limit for $A_V > 0.3$ mag.

Actually, both SN~1991T and \bu\/ show evidence of significant
extinction from interstellar dust associated with the host galaxy,
$A_V^{host} = 0.46$ for SN~1991T and $A_V^{host} = 0.86$ for \bu\/
(Tab.~1).  These values were derived mainly by photometric methods
based on the analysis of the light curve and the SN color, and
therefore they are affected by large uncertainties.  In a simplified
scenario, one can assume that there is only one cloud which is causing
both the observed extinction and the light echo. In this case, since
we fix the column density of the cloud, we can derive the distance $D$
of the cloud itself from the SN through a fit to the echo luminosity.
The light curve fit for SN~1991T, shown in Fig.~1, gives $D\sim 120$
lyr.  In the case of SN~1998bu the light curve is similar to that of
SN~1991T, hence $\Delta t_{SN} \simeq 0.1$ yr, but the optical depth
of the dust is about a factor 2 larger: therefore we obtain $D\sim
230$ lyr.  Note that these estimates are independent of the adopted
distance for the parent galaxies, because they depend only on the
ratio between the luminosity of the echo and that of the SN at
maximum.

{
\centerline{{\vbox{\epsfxsize=8cm\epsfbox{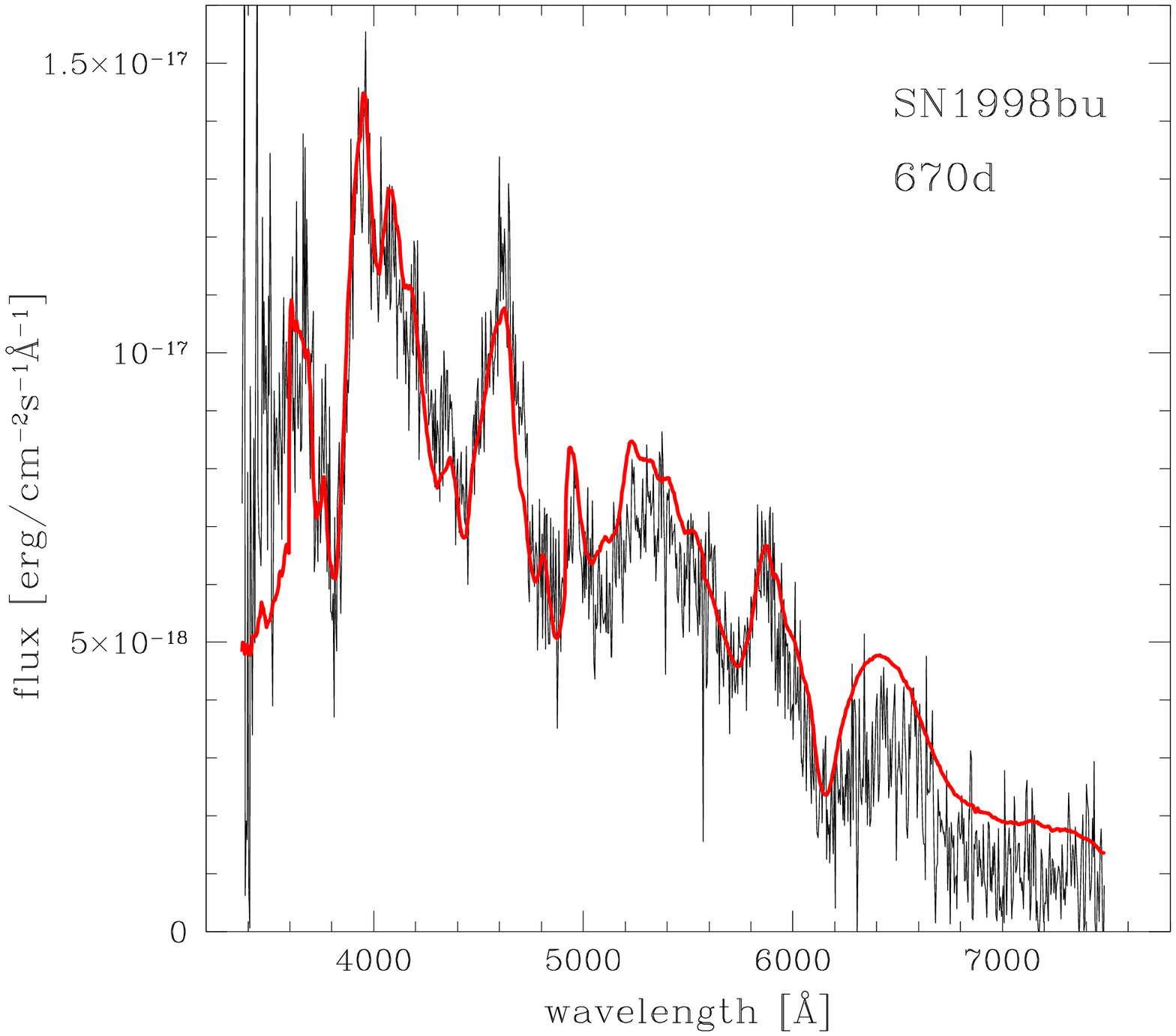}}}}
{\footnotesize{Comparison of the observed spectrum of SN1998bu at phase 670d 
(black line) and with the time integrated echo spectrum computed as
described in the text (red line).}}
\bigskip}

The distance $D$ is related to the linear diameter of the ring through
the relation $2R = 2\sqrt{ct\;(2D+ ct)}$. In the case of SN~1991T the
echo has been resolved with a diameter of 0.$^{\prime\prime}$4.  Based
on the Cepheid distance to NGC~4527, this results in a linear diameter
$2R = 120$ lyr and hence $D \simeq 150$ lyr.  Given the observational
uncertainties, this is in very good agreement with the estimate based
on the light echo modeling.  In the case of SN~1998bu, because of the
much higher optical depth of the dust, we had to place the dust layer
at a larger distance than for SN~1991T.  In turn, this causes the echo
ring diameter at any given time to be larger. Based on the geometry
described earlier we obtain that at the time of writing, 2.5 yr after
the explosion, the diameter of the echo should be $\sim 0.3$
arcsec. Indeed, the echo of SN~1998bu has already been resolved using
HST (Leibundgut p.c.).

A consistency check of the light echo interpretation for the peculiar
light curve of \bu\/ can be obtained by comparing the observed and
computed spectrum of the light echo.  Clearly the most important
contribution to the light echo comes from the light emitted near
maximum. Spectra of \bu\/ at eight different epochs from -6 to 50 days
after maximum were published by \citet{hernandez}. Weighting the
spectra according to the integrated luminosity around each observed
epoch and using our assumed geometry (cf. Equation 1), we computed the
contribution of each early-time spectrum to the emerging echo spectrum
at phase 670d. The input spectra were corrected for reddening using a
standard extinction law and assuming that the scattering function as a
similar $\lambda^{-1}$ dependence.  These corrected spectra were
coadded to compute the expected spectrum of the echo (at 670d the SN
nebular spectrum is several magnitudes fainter than the echo and does
not give a measurable contribution). This is compared with the
observed spectrum in Fig.~3. Given the crudeness of some of the
assumptions and the incomplete spectral coverage of the near-maximum
phase, the agreement between the observations and the model is
excellent, which we consider a strong argument in favor of the light
echo scenario.

Moreover, even if we stress that similar results can be obtained
for different geometries, the fact that with the configuration we have
chosen it is possible to explain the observed light curve, spectrum
and reddening is suggestive.

\section{Are bright SN~Ia linked to dusty star forming regions ?}

The observation and interpretation of light echos can 
tell us about dust properties and distribution in galaxies, 
but may have far-reaching applications and consequences. 
One such application is to measure distances of galaxies \citet{sparks}.

SN~1998bu is only the second Type Ia SN for which a light echo was
observed, the other being SN~1991T in the Sbc galaxy
NGC~4527. It is remarkable that both SNe are
very slow decliners. SN~1991T was also spectroscopically peculiar at
and before maximum, indicating a higher photospheric temperature than
in normal SNe~Ia, as shown by the presence of strong FeIII lines
and by the absence of FeII lines before maximum \citep{filip}. This
was confirmed by spectroscopic modeling, which also revealed that
SN~1991T had an abnormally high abundance of Fe-group species in the
outer, fast-moving part of the ejecta \citep{paolo95}. Such a peculiar
element distribution suggested that SN~1991T may have been the result
of an unusual explosion mechanism.  Further studies revealed that
SN~1991T had very broad Fe nebular lines \citet{paolo98}. This and
the brightness of the SN at maximum led various authors to conclude
that SN~1991T produced an unusually large amount of $^{56}$Ni, about 
1 \Msun \citep{spyro,paolo95,cap97,paolo98} though this may be
challenged by recent calibration of the SN absolute luminosity
\citep{saha,richtler,gibson}. Finally, \citet{fisher} suggested that
SN~1991T may have been the result of the explosion of a
super-Chandrasekhar mass progenitor.

SN~1998bu shares many of the properties of SN~1991T: it was bright at
maximum, it had a slow decline and broad nebular lines, although not
quite as extreme in these respect as SN~1991T.  It is therefore very
interesting that both SNe are significantly reddened and show a dust
echo.  Other two nearby SNe~Ia were heavily reddened: SN~1989B was a
normal SN~Ia for which indeed there may be some evidence of a light
echo \citep{milne}, while SN~1986G was a fast decliner and
spectroscopically peculiar object for which, despite an early claim
\citep{schaefer}, there was no evidence of a light echo
\citep{turatto}.  The latter requires that the dust was more distant
from SN~1986G than in \bu\/ or SN~1991T (by placing the dust
cloud 10 times further the echo magnitude becomes 2.5 mag fainter).

Historically, the fact that SN~Ia were seen as very homogeneous was
attributed to their progenitors arising from a single stellar
population. Adding to this that SN~Ia were found in all types of
galaxies, even in Ellipticals, contributed to the standard paradigm that
the progenitors of all SN~Ia belong to the old stellar population.
Now there is evidence that slow decliners (or high luminosity SN~Ia)
may be preferably associated with a younger stellar population (and
therefore conceivably more massive progenitors), a suggestion first made by 
\citet{vdb} and confirmed by \citet{hamuy,hamuynew}.
Our result that both SN~1991T  and
SN~1998bu are located close to dusty regions is coherent with this
scenario.

Obviously one needs to improve the statistics by adding more
cases. In this respect one interesting object is SN~1998es.  This is a
SN~Ia whose spectrum near maximum was very similar to that of SN~1991T
\citep{jha98} and shows the signature of quite similar interstellar reddening
(Patat et al. in preparation). Therefore SN~1998es
 maybe an interesting candidate to search
for light echo though, because of the relatively large distance ($\mu
= 33.2$), the apparent magnitude may result quite faint (V$\sim 24$ if
the echo has the same intensity as in SN~1991T and \bu\/).

\section{Conclusions}

The peculiar late light curve and spectrum of SN~1998bu are attributed
to the echo from circumstellar dust of the light emitted near maximum.
Based on a simple modelling and on the comparison with SN~1991T (which
also experienced a similar phenomenom) we estimate that the dust cloud
is located in front of the SN and relatively nearby ($\sim 100
pc$).  Because of the similar echo luminosity but much higher optical
depth of the dust in \bu\/ compared with SN~1991T, we expect that the
dust layer is more distant and hence the echo ring of \bu\/ grows faster
than in SN~1991T.  

The association of dust with SN~Ia, possibly of a specific subtype,
may have interesting implications for the progenitor scenario and
prompts for a renewed effort for monitoring SN~Ia at late phases.

\end{document}